
\magnification = \magstep 1
\hsize = 32 pc
\vsize = 42 pc
\baselineskip = 24 true pt
\def\vs{\vskip 0.5 true cm}
\def\nd{\noindent}

\nd\hfill {IP-BBSR /93-64} \break

\nd\hfill {October-93} \break
\vs
\vs
\nd\centerline{\bf Some exactly solvable three-body problems}
\nd\centerline{\bf in one dimension}
\vs
\vs
\nd\centerline{\bf Avinash Khare}

\nd\centerline{Institute of Physics,Sachivalaya Marg}

\nd\centerline{Bhubaneswar 751005,India.}

\vs

\nd\centerline{\bf Rajat K. Bhaduri}

\nd\centerline{Department of Physics and Astronomy, McMaster University}

\nd\centerline{Hamilton, Ontario,Canada, L8S 4M1.}
\vs
\vs
\nd{\bf Short Title: Three-body problem}

\nd{\bf PACS 03.65.Ge, 21.45.+v }
\vs
\nd{\bf Abstract}

   The three-body problem in one-dimension with a repulsive inverse
square potential between every pair was solved by Calogero. Here, the
known results of supersymmetric quantum mechanics are used to propose
a number of new three-body potentials which can be solved algebraically.
Analytic expressions for the eigenspectrum and the eigenfunctions are
given with and without confinement.

\vfill

\eject

\nd{\bf I. Introduction}

Long back, in a classic paper (Calogero 1969), Calogero had given the
complete solution of the Schr\"odinger equation for three particles in
one dimension, interacting pairwise by two-body harmonic and
inverse-square potentials. Later, Wolfes (Wolfes 1974) had used Calogero's
method to obtain analytical solutions of the same problem in the
presence of an added three-body potential of a special form.
Attention soon shifted to the exact solutions of the many-body problem
 (Sutherland 1971,Calogero 1971) and the general question of integrability.
A list of solvable pair potentials for the many-body problem is given
in the review by Olshanetsky and Perelomov (Olshanetsky and Perelomov
 1981, 1983),where other references will be found. Recently, there has been a
renewed interest in the one-dimensional many-body problem of the Calogero
and the Sutherland types (Polychronakos 1992), and their applications to
the physics of spin chains (Haldane,1988, Shastry 1988, Frahm 1993).In
this paper, we give some new potentials that have exact solutions of the
three-body problem in one-dimension. Examples of confined spectra of
the Calogero-Wolfes type, as well as other cases with both bound and
scattering states present, are given.

     Three particles in one dimension, after the
cetre-of-mass motion is eliminated, have two independent degrees of
freedom. This may therefore be mapped on to a one-body problem in two-
space dimensions as was done by Calogero. For the type of potentials used,
the radial and angular parts separated, and the eigenvalue
problem rested essentially on the availability of an analytical solution
for the angular part. The same method is followed here, but advantage
is also taken of supersymmetric quantum mechanics . The examples with
solutions in a closed form that are presented here are found to be shape
invariant supersymmetric potentials (Gendenshtein 1983,  Dutt et al
 1988, Khare et al 1991) in the angular variable.
Bound state algebraic solutions with harmonic confinement are given in Section
II, while scattering is discussed in Section III. The harmonic confinement of
Section
II may be replaced by a ${1\over{r}}$-type potential, and this is briefly
discussed in the final Section.
\vs
\nd{\bf II. New examples with harmonic confinement}

   Calogero had given the solution for the following potential for the
three-body problem :
$$
  V_{C}= \omega^2/8 \sum_{i<j} (x_{i}-x_{j})^2 + g \sum_{i<j}
(x_{i}-x_{j})^{-2}
 \eqno(1)$$
where $g>-1/2$ to avoid a collapse of the system.Wolfes (Wolfes 1974)
showed that a three-body potential
$$
V_{W}= f[(x_{1}+x_{2}-2x_{3})^{-2}+(x_{2}+x_{3}-2x_{1})^{-2}
       +(x_{3}+x_{1}-2x_{2})^{-2}]
\eqno(2) $$
is also solvable when it is added to $V_{C}$, with or without the
pair-wise centrifugal term.The last two terms on the right-hand side of Eq.(2)
are just cyclic permutations of the first. Henceforth, such terms occuring in
other potentials would be referred to as 'cyclic terms '. In this section, we
give more examples
of three-body potentials that may be solved exactly. Our first solvable
example is the three-body potential of the form
$$
V_{1}={{\sqrt{3}f_{1}}\over{2
r^{2}}}[{{(x_{1}+x_{2}-2x_{3})}\over{(x_{1}-x_{2})}}+\; \rm {cyclic\;
terms}\;]\;,
\eqno(3) $$
which is added to $V_{C}$ of Eq.(1).
In the above equation,
$$
    r^2={{1}\over{3}} [(x_{1}-x_{2})^2+(x_{2}-x_{3})^2+(x_{3}-x_{1})^2].
\eqno(4) $$
To see why potentials given by Eqs(1)-(3) are solvable, define the
Jacobi coordinates

$$    R={{1}\over{3}}(x_{1}+x_{2}+x_{3}),$$
$$    x={{(x_{1}-x_{2})}\over{\sqrt{2}}},\;\;
y={{(x_{1}+x_{2}-2x_{3})}\over{\sqrt{6}}}.\eqno(5)$$
After elimination of the centre-of-mass part from the Hamiltonian, only
the x- and the y-degrees of freedom remain, which may be mapped onto the
polar coordinates
$$
    x=r\,\rm{sin}\,\phi,\;\; y=r\,\rm{cos}\,\phi,\eqno(6) $$
where we follow the notation of Calogero.Obviously, the variables $r$, $\phi$
 have ranges  $0\leq r\leq\infty$, and $ 0\leq \phi\leq2\pi$.It is straight-
forward to show that

$$ (x_{1}-x_{2})=\sqrt{2}\;r\;\rm{sin}\,\phi,$$
$$ (x_{2}-x_{3})=\sqrt{2}\;r\;\rm{sin}(\phi+
 2\pi/3),$$
$$ (x_{3}-x_{1})=\sqrt{2}\;r\;\;\rm{ sin}(\phi+4\pi/3).\eqno(7)$$
The Schr\"odinger equation separates cleanly in the radial and angular
variables, and the wave function may be written as
$$
        \Psi_{nl}(r,\phi)= R_{nl}(r)F_{l}(\phi).
\eqno(8) $$
The radial wave function obeys the equation $(\hbar=1,2M=1)$.
$$
 ({{ -d^2}\over{dr^2}}-{{1}\over{r}}{{ d}\over{dr}}+{{3}\over{8}}
w^2r^2+B_{l}^2/r^2)R_{nl}=E_{nl}R_{nl}(r),
\eqno(9)$$
where $B_{l}^2$is the eigenvalue of the Schr\"odinger equation in the angular
 variable. The eigenvalues $E_{nl}$
 are given by
$$
 E_{nl}=\sqrt{3/2}\;\omega(2n+B_{l}+1).\;\; n=0,1,2...;\;\; l=0,1,2...
\eqno(10)$$

The form of $R_{nl}$ is given by (Calogero 1969)
$$
 R_{nl}=(r)^{B_{l}}
\exp[{{-1}\over{4}}(\sqrt{3/2})\omega
r^2]L_{n}^{B_{l}}[{{1}\over{2}}(\sqrt{3/2})\omega r^2],
\eqno(11)$$

where$ B_{l}>0$.To examine the angular part of the eigenfunction
$F_{l}(\phi)$,take the potential $V_{C}+V_{1}$.Then, in the variable $\phi$,
the
Schr\"odinger equation is
$$[{{-d^2}\over{d\phi^2}}+{{g}\over{2}}(\sum_{m=1}^{3}{\rm cosec}^2[\phi+2(m-1)
{\pi\over{3}}])+{{3}\over{2}}\,f_{1}(\sum_{m=1}^{3}{\rm
cot}[\phi+2(m-1){\pi\over{3}}])]F_{l}(\phi)=B_{l}^2\,F_{l}(\phi).\eqno(12)$$
Calogero used the identity
$$ \sum_{m=1}^{3}{\rm cosec}^2[\phi+2(m-1)\pi/3]=9{\rm
cosec}^2(3\phi).\eqno(13)$$
A similar identity may be used for ${\rm cot}\,\phi$,
$$\sum_{m=1}^{3}{\rm cot}[\phi+2(m-1)\pi/3]=3{\rm cot}(3\phi).\eqno(14)$$
Using Eqs(13) and (14), (12) reduces to
$$[-{{d^2}\over{d\phi^2}}+{{9}\over{2}}\,g\,{\rm cosec}^2(3\phi)+{{9}\over{2}}
\,f_{1}\,{\rm cot}(3\phi)]\,F_{l}=
 B_{l}^2\,F_{l}(\phi),\eqno(15)$$
whose solution will be given presently.

      The solution of Eq(15) and some other examples to be given are found by
algebraic methods of supersymmetric quantum mechanics (Infeld and Hull 1951,
Gendenshtein 1983). Define a ``super
potential'' $W(\alpha\pi)$, where $\alpha$ is a constant and $\phi$ the
angular variable. Then the supersymmetric partners $V_{-}$ and $ V_{+}$ are
$(W^2-W')$ and $(W^2+W')$ respectively,the prime denoting a differentiation
with respect to $\phi$.The potential $V_{-}$ has a zero energy ground state
with the wave function $F_{0}$ given by $\exp(-\int_{ }^{\phi}W d\phi )$.
In the literature (Dutt et al 1987, Levai 1989, Khare et al 1991 ), various
superpotentials $W$ are given for which $V_{-},V_{+}$ are exactly solvable.
In these examples,$ V_{-}$ and $V_{+}$ are ``shape invariant'', meaning
thereby that $V_{+}$ is identical in functional form to $V_{-}$ with an
appropriate change in one or more parameters. The potential in Eq(15) is
obtained if we choose
$$ W = -A{\rm cot}\,{3\phi}-C/A,\;\;\; (A>0),\eqno(16)$$
where C,A are parameters independent of $\phi$.Then
$$V_\mp= A(A\mp3){\rm cosec}^2{3\phi}+2C{\rm
cot}\,{3\phi}+(C^2/A^2-A^2).\eqno(17)$$
In this example, $ V_{\pm}$ are shape invariant potentials,i.e. they are
related as
$$V_{+}(A,C,\phi)=
V_{-}(A+3,C,\phi)+(A+3)^2-C^2/(A+3)^2+C^2/A^2-A^2.\eqno(18)$$
This in turn implies that the energy eigenvalues of $V_{-}$
are given by
$$B_{l}^2= (A+3l)^2-C^2/(A+3l)^2-A^2+C^2/A^2.\eqno(19)$$
 From this, it immediately follows that for the potential
$$V={{ 9}\over{2}} g\;{\rm cosec}^2{3\phi}+{{9}\over{2}}\;f_{1}{\rm
cot}\,{3\phi}\eqno(20)$$
occuring in Eq(15), the eigenvalue spectrum is
$$B_{l}^2= 9(l+a+1/2)^2-{{9}\over{16}} f_{1}^2/(l+a+1/2)^2,\eqno(21)$$
where
$$a= 1/2(1+2g)^{1/2}\;.\eqno(22)$$
As in Calogero, $g>-1/2$ for meaningful solutions.The unnormalized ground state
wave function is given by
$$F_{0}(\phi)= \exp[-\int^{\phi}Wd\phi]=({\rm sin\,
3\phi})^{a+{1\over{2}}}\;\exp[{{3}\over{4}}f_{1}\phi/(a+1/2)],\eqno(23)$$
valid in the range$ 0\leq3\phi\leq\pi.$
For distiguishable particles, a given value of $\phi$ defines a specific
ordering. For ${0\leq\phi\leq\pi/3}$, Eq(7) implies $x_{1}\geq x_{2}\geq
x_{3}$,and other ranges of $\phi$ correspond to different orderings (Calogero
1969,
Wolfes 1974).For singular repulsive potentials, crossing is not allowed,
and $F_{0}(\phi)$ of Eq.(23) is zero outside $0\leq\phi\leq\pi/3$. Following
Calagero, the wave function for the other ranges may be constructed.For
indistiguishable particles,similarly, symmetrised or antisymmetrised wave
functions may be constructed  and this will not be
repeated here. From the table given by Levai (Levai 1989) the general solution
$F_{l}(\phi)$ of Eq.(15) is given by
$$F_{l}(\phi)=\exp(-i{\pi\over 2}{\tilde l})({\rm sin}\,3\phi)^{\tilde l}
 \exp[{3\over4}{{ f_{1}\phi}\over{\tilde l}}] P_{l}^{-{\tilde l}
-{{if_{1}}\over{4{\tilde l}}},-{\tilde l}+{{if_{1}}\over{4  {\tilde l}}}}
(i\;\rm{cot}\,3\phi).\eqno(24)$$
In the above equation, $ P^{\alpha,\beta}_{n}$ is the Jacobi polynomial of the
arguement$(i\rm cot{3\phi}).$
Another exactly solvable potential is given by
$$\eqalignno{V_{2}&={{1}\over{8}} \omega^{2}\sum_{i<j}(x_{i}-x_{j})^2\cr
&\;+3 g\; [(x_{1}+x_{2}-2x_{3})^{-2}+\;\rm{cyclic\;terms}\;]\cr
& -{{3\sqrt{3}}\over{2}}\;{{
f_{1}}\over{r^2}}\;[{{(x_{1}-x_{2})}\over{(x_{1}+x_{2}-2x_{3})}}+\;\rm{cyclic\;terms}\,]\;.&(25)\cr}$$
Although this looks very different from the previous example of
$(V_{C}+V_{1}),$ it is simply obtained by the transformation$
\phi\rightarrow\phi+\pi/2$. A little
algebra shows that Eq.(25) reduces to
$$V_{2}={{3}\over{8}}{ \omega^2} r^2 +{{g}\over{
2 r^2}}{ \sum_{m=1}^{3}}{\rm sec}^2 [\phi+
{{2(m-1)\pi}\over{3}}]-{{3}\over{2}}{{ f_{1}}\over{r^2}}{ \sum_{m=1}^{3}}{\rm
tan}[\phi+{{2(m-1)\pi}\over{3}}]$$
$$       ={{3}\over{8}} \omega^2 r^2 +9{{g}\over{2 r^2}}{{ \rm
sec}^2{3\phi}}-9{{ f_{1}}\over{2 r^2}}{\rm tan}\,{3\phi}\;.\eqno(26)$$
The angular part of $V_{2}$ is identical to Eq.(20) if $\phi\rightarrow \phi +
\pi/2$.
Due to this shift, the allowed range of $\phi$ will divide into sectors
$- \pi/6\leq \phi\leq \pi/6,\,\pi/6\leq \phi\leq \pi/2,$ etc (Wolfes
1974).Otherwise,
the solution $F_{l}(\phi)$ is the same as Eq.(24) with $\phi\rightarrow
\phi+\pi/2,$
with $ B_{l}^2$ still given by Eq.(21).
There are more examples to be found by exploiting supersymmetric quantum
mechanics. A potential that is exactly solvable, but is different from the
example given by Wolfes (Wolfes 1974) is
$$V_{3}={{1}\over{8}} \omega^2\sum_{i<j} (x_{i}-x_{j})^2
+g\sum_{i<j}(x_{i}-x_{j})^{-2}-{{f_{3}}\over{\sqrt{6} r}}[{{(x_{1}+x_{2}-2
x_{3})}\over{(x_{1}-x_{2})^2\,}}+
\,\rm{cyclic\;terms}\,].\eqno(27)$$
Using relations (7) and (13), and the identity,
$$\sum_{m=1}^{3}\rm cos(\phi+2(m-1)\pi/3)/\rm sin^2(\phi+2(m-1)\pi/3)=
9\rm cos\, {3\phi}/\rm sin^2{3\phi},\eqno(28)$$
$V_{3}$ reduces to the form
$$V_{3}={{3}\over{8}} \omega^2 r^2 +{{9}\over{2}}{{g}\over{r^2}}{\rm cosec}^2
{3\phi}-{{9}\over{2}}{{f_{3}}\over{r^2}}{\rm cot}\,{3\phi}\,{\rm
cosec}\,{3\phi}.\eqno(29)$$
As before, the Schr\"odinger equation is separable in $r$ and $\phi$, with the
radial solution being the same as in Eq.(11), and the angular part reduces  to
$$[-{{d^2}\over{{d\phi^2}}}+{{9}\over{2}}\,g\,{\rm
cosec}^2{3\phi}-{{9}\over{2}}\, f_{3}\,{\rm cot}\,{3\phi}\,
{\rm cosec}\,{3\phi}]F_{l}(\phi) = B_{l}^2\, F_{l}(\phi).\eqno(30)$$
This form of $V_{-}(3\phi)$ is obtained by taking the superpotential ( Dutt
etal.\ 1988 )$W$ to be
$$W=-A\;{\rm cot}\,3\phi + C\;{\rm cosec}\, 3\phi\eqno(31),$$
with $0\leq 3\phi\leq\pi,\;\; A > 0$, and $C < A$.The resulting $V_{-}$
is then given by
$$V_{-}=(A^2+C^2-3A){\rm cosec}^2 {3\phi}-(2A-3)C\,{\rm
cot}\,{3\phi}\,{\rm\,cosec}3\phi -
A^2,\eqno(32)$$
which is the same as in Eq (30) apart from the constant$ A^2 $ .
The eigenvalues $B_{l}^2$ are then independent of $C$, and given by
$$B_{l}^2 = ( A + 3 l )^2 ,\;\;\; l = 0,1,2,.\;\;   .\eqno(33)$$
These are of the same form as obtained for the pair-wise inverse square
potential.The unnormalised angular wave function  $F_{l}(\phi)$ is given by
$$F_{l}(\phi)= ({\rm sin}\,{{3\phi}\over{2}})^{\alpha}({\rm
cos}\,{{3\phi}\over{2}})^{\beta}P_{l}^{\beta-1/2,\alpha-1/2}\,\,\,({\rm
cos}\,3\phi)\;\;,\eqno(34)$$
where $\alpha = (A+B)/3,\; \beta = (A-B)/3 $. Note that for the wave function
$ F_{l}$ to vanish at $3\phi = 0 $ and at $3\phi = \pi,\;$$\alpha$ and $\beta
$ should both be $>0$.The range of the variable $\phi$ may be extended
following Calogero ( Calogero 1969 ).
A different form of a solvable potential may again be obtained from Eq.(29)
by transforming $\phi\rightarrow\phi+\pi/2.$ From Eq.(6), this changes
$x\rightarrow y$ and $y\rightarrow -\;x.$The resulting form of the solvable
potential  is given by
$$V_{4}=\,{{1}\over{8}}\omega^2\sum_{i<j}(x_{i}-x_{j})^2\,+3\,g\,[\,(x_{1}+x_{2}-x_{3})^{-2}\,+\;\rm{cyclic\;terms}\,]\,$$
$$+{1\over{3\sqrt{2}}}{f_{3}\over r}\,[\,{{(x_{1}-
x_{2})}\over{(x_{1}+x_{2}-2x_{3})^2}}+\;\rm{ cyclic\;terms}\,].\eqno(35)$$
The reader may obtain the eigenvalues and the eigenfunctions of this potential
 (valid for the sectors
$-\pi/6\leq\phi\leq\pi/6,\;\;\pi/6\leq\phi\leq\pi/2$ etc.) from the solutions
of $V_{3}$, and the details would not be given here.

There are other types of three-body potentials that may be constructed whose
angular part is solvable, but is not of the supersymmetric shape-invariant form
as before.In such examples, however, although the eigenfunctions $F_{l}(\phi)$
are expressible in terms of well-known functions, the eigenspectrum of
$B_{l}^2$ is not found to be in a simple closed form.For example,keeping in
mind the
identity,
$$-\,(x_{1}-x_{2})(x_{2}-x_{3})(x_{3}-x_{1})={1\over{\sqrt{2}}}\,r^3\,\rm{sin}\,3\phi,\eqno(36)$$
a three-body potential of the form ${1\over{r^2}}\,\rm{sin}^2{3\phi}$ may be
constructed. Combining this with the pair
potential of Calogero that yielded $\,1/(r^2\,\rm{sin}^2{3\phi})$, the angular
part of the three-body Schr\"odinger equation may be written in the form
$${{d^2F_{l}}\over{d\phi^2}}+[B_{l}^2\,-\,a\,{\rm cos}\,6\phi\,-\,{b\over{{\rm
sin}^2\,3\phi}}]\,F_{l}\,=\,0\,.\eqno(37)$$
It is not difficult to show that this may be reduced to the well-known
differential equation (Erdelyi 1955) for spheroidal wave functions ( for
$b>a-1/2$ ), whose eigenvalues, even though known, are not expressible in a
simple closed form.

Till now, we have been discussing only those cases where the Schr\"odinger
equation in the variables $r$ and $\phi$ are separable, and exactly solvable.
The
three-body potential given by Eq.(36) is of special interest, however, even
though its form does not allow separation of the coordinates $r$ and $\phi$.
To be specific, consider the potential
$$V_{H}=\,{1\over{8}}\omega^2\,\sum_{i<j}\,(x_{i}-x_{j})^2\,-\,\lambda\,(x_{i}-
x_{2})(x_{2}-x_{3})(x_{3}-x_{1}).\eqno(38)$$
Transforming to polar coordinates using Eqs (7), this reduces to
$$V_{H}=\,{3\over{8}}\,\omega^2\,r^2\,+\,{\lambda\over{\sqrt{2}}}\,r^3\,{\rm
sin}\,3\phi,\;\eqno(39)$$
which is the famous  Henon-Heiles potential (Henon and Heiles 1964 ) for a
particle in 2-space dimensions. This potential is nonintegrable, and has been
extensively studied in connection with chaos. Therefore $V_{H}$ given by
Eq(38) may be regarded as an example of a potential that gives rise to
classical chaotic motion of three particles in one spatial dimension. Since it
is
well-known that classical periodic orbits have a close connection with the
quantum density of states (Gutzwiller 1990), it will be of interest to map
the two-dimensional periodic orbits of the Henon-Heiles potential (Brack
 et al.\ 1993) onto the one dimensional motion of the three particles.
\vs
\nd{\bf III. The three-body scattering problem}

   The three-body scattering problem with the inverse square pair potential,
$$g\,\sum_{i<j}\,(x_{i}-x_{j})^{-2},$$
was solved by Marchioro (Marchioro 1969).
When an additional three-body potential $V_{W}$ (see Eq.(2)) is added to
this, the problem is still solvable (Calogero and Marchioro 1974).Both the
classical and quantum solutions are strikingly simple in these problems.
Asymptotically, with the inverse square pair potential, the momenta of the
scattered particles, $p_{i}^{'}$ are related to the incident ones $p_{i}$ by
$p_{i}^{'}=\,p_{4-i}\;,\,(i=\,1,2,3\,)$, and the total phase shift is
independent of $l$. With the addition of the three-body potential $V_{W},$ the
asymptotic momenta just change sign, and the phase shift still remains
$l$-independent.We
have also studied the scattering problem with the new potentials (cited in
the last section ) after dropping the harmonic confinement. Even though the
problems are still integrable, we find that the simplicity of the Calogero -
 Marchioro examples is lost. The scatterings with the new interactions
cannot simply be expressed as an exchange between the incoming and asymptotic
outgoing momenta.
As an example, consider the classical scattering of three particles with the
potential
$${\tilde V_{3}}=\,g\,\sum_{i<j}\,(x_{i}-x_{j})^{-2}\,-\,{1\over{\sqrt{6}}}
{{f_{3}\over{r}}}[{{(x_{1}+x_{2}-2x_{3})}\over{(x_{1}-x_{2})^2}}+\,\rm{cyclic
\;terms}\;],\eqno(40)$$
which is the same as $V_{3}$ given by Eq.(27), but without the harmonic
confinement.Because the variables $r,\,\phi$ are separable, we may write,
following Marchioro's notation ( Marchioro 1969 ),
$${1\over{2}}\,p_{r}^2+{{B^2}\over{r^2}}\,=\,E\,,\eqno(41)$$
$${1\over{2}}\,p_{\phi}^2+{9\over{2}}\,g\,{\rm cosec}^2\,3\phi\,-\,{9\over{2}}
\,f_{3}\,{\rm cot}\,3\phi\,{\rm cosec}\,3\phi\,=\,B^2\,.\eqno(42)$$
Here $E$ is the total energy and $B$ the angular constant of motion.It is
straight-forward to perform the integrations. Writing
$$b\,={{9\,f_{3}}\over{2\,B^2}}\,,\;\;\;c={{18\,g}\over{B^2}}\,,\eqno(43)$$
we obtain
$$r(t)\,=\,[2\,E\,(t-t_{0})^2\,+\,{{B^2}\over{E}}]^{{1\over{2}}}\,,\eqno(44)$$
and
$${\rm cos}\,3\phi\,=\,{b\over{2}}\,-\,{1\over{2}}\sqrt{4-c+b^2}
\,{\rm sin}\;[{\rm sin}^{-1}(\,{{b-2\,{\rm
cos}\,3\phi_{0}}\over{\sqrt{4-c+b^2}}})
\,+3\,{\rm tan}^{-1}(\sqrt{2}\,{E\over{B}}\,(t-t_{0}))]\,.\eqno(45)$$
In the above, $\phi=\,\phi_{0}$ at $t=\,t_{0}$.Let $\phi=\,\phi_{i}$ for
$\,(t-t_{0})\rightarrow\infty,$ and $\phi=\,\phi_{f}$ for
$(t-t_{0})\rightarrow\infty.$Then

$${\rm cos}\,3\phi_{i}=\,{b\over{2}}-\,{1\over{2}}\sqrt{4-c+b^2}\,
{\rm cos}[\,{\rm sin}^{-1}({{b-2\,{\rm
cos}\,3\phi_{0}}\over{\sqrt{4-c+b^2}}})]\,,$$
and
$${\rm cos}\,3\phi_{f}=\,{b\over{2}}+\,{1\over{2}}\,\sqrt{4-c+b^2}\,
{\rm cos}[\,{\rm sin}^{-1}({{b-2\,{\rm
cos}\,3\phi_{0}}\over{\sqrt{4-c+b^2}}})]\,.\eqno(46)$$
Therefore only for $b=0$,i.e.,for $f_{3}=0$ ,does ${\rm cos}\,3\phi_{i}=
-{\rm cos}\,3\phi_{f}$, leading to $3\phi_{i}=\,(\pm\,3\phi_{f}+\pi).$
If one considers the sector for which $0\leq\phi\leq\pi/3 $, then
$\phi_{i}=\,-\phi_{f}+\pi/3$ for $b=0$. It was this peculiarity that yielded
the simple relationships between the initial and the final asymptotic
momenta of the particles. But for $f_{3}\neq 0 $, no such relationship exists,
and the simplicity in the scattering process is lost.
For the same example, similar complications arise in the quantum treatment
of the scattering problem with $f_{3}\neq 0$. In this case the angular wave
function given by Eq.(34) does not possess a simple symmetry property for
$\phi\rightarrow \phi-\pi/3$ when $\alpha\neq \beta.$ For other examples
like $V_{1}$(given by Eq.(3)) plus the inverse square pair potential, the
situation is even more complicated, and will not be discussed further
\vs
\nd{\bf IV. Discussion}

   The shape-invariant potentials in the angular variable $\phi$ were combined
with a harmonic confinement to obtain exact solutions in Section 2. This led
to a discrete eigenvalue spectrum. The harmonic confinement may be replaced
by an attractive ${1\over{r}}$-type interaction,giving rise to both discrete
and continuous energies. Exact solutions for all the shape-invariant potentials
described earlier can again be obtained algebraically. As one example, take the
potential
$$\eqalignno{V&=\,-{{\sqrt{3}\,\alpha}\over{\sqrt{(x_{1}-x_{2})^2+(x_{2}-x_{3})^2+(x_{3}-x_{1})^2}}}+\,g\sum_{i<j}(x_{i}-x_{j})^{-2}\;,\cr
&=\,-{\alpha\over{r}}\,+\,{9\over{2}}\,\,{{g\over{r^2\,{\rm
sin}^2\,3\phi}}}\;.&(47)\cr}$$
First consider the classical problem.Using the notation of Eqs. (41) and (42),
we obtain,
$${{2\,B^2}\over{\alpha\,r}}=\,1+\sqrt{1+{{4\,E\,B^2}\over{\alpha^2}}}
{\rm cos}[{1\over{3}}\,{\rm cos}^{-1}({{\rm{cos}\,3\phi}\over{\sqrt{1-9\,g/
(2\,B^{2})}}})]\,.\eqno(48)$$
For $g=0$, and $E<0$, Eq.(48) reduces to the periodic orbit of an ellipse in
the polar coordinates :
$${p\over{r}}=\,1+\,e\;\rm{cos}\,\phi\,,\eqno(49)$$
where $p=\,{{2B^2}\over{\alpha}}$is half the latus rectum, and
$e=\,\sqrt{1-{{4|E|B^2}\over{\alpha^2}}}$ is the eccentricity of the
ellipse(Landau and Lifshitz 1960).For $g=0$,crossing between any pair of
particles is allowed, and the periodic orbit given by Eq.(49) may be mapped
on to the motion of three particles along a line. On the other hand, for
$g\neq 0$,Eq.(48) does not reduce to a closed orbit for the bound problem.
For the special case when $g\neq 0$ between one pair, and $0$ between the
other two pairs, Eq.(48) still reduces to the closed orbit form of Eq.(49).
However, crossing between the interacting pair is not allowed,and the mapping
in each sector has to be done accordingly.
The Schr\"odinger equation with the potential $V$ of Eq.(47) is easy to solve.
Writing the wave function $\psi_{nl}(r,\phi)=\,R_{nlo}(r)F_{l}(\phi)$ as
before, the radial part obeys the standard Coulomb-type equation with the
bound-state wave function
$$R_{nl}(r)=\,r^{B_{l}}\,\exp(\,-\sqrt{|E_{nl}|}\,r)\,L_{n}^{2B_{l}}(2r/r_{0})
\,,\eqno(50)$$
where $r_{0}=2(B_{l}+n+{1\over{2}})/\alpha,$ and $L_{n}^{2B_{l}}$ is
the Laguerre polynomial.The corresponding eigenvalues are
$$E_{nl}=\,-{{\alpha^2}\over{4(n+B_{l}+{1\over{2}})^2}}\,,\;n=0,1,2,..
\;\;l=0,1,2..\,.\eqno(51)$$
As before, the angular constants $B_{l}$'s are the eigenvalues of the equation
$$-{{d^{2}}\over{d\phi^{2}}}+\,{9\over{2}}{g\over{\rm{sin}^2\,3\phi}}\,F_{l}(\phi)=\,B_{l}^2\,F_{l}(\phi).\eqno(52)$$
The unnormalised solutions$ F_{l}$'s are expressed in terms of the
Gegenbauer polynomials (Calogero 1969)
$$F_{l}=\,({\rm sin}3\phi)^{a+{1\over{2}}}\,C_{l}^{a+{1\over{2}}}({\rm cos}3
\phi)\;,\;\;\;0\leq\phi\leq {\pi\over{3}}\;.\eqno(53)$$
Here the constant $a$ is given by Eq.(22), and the wave function vanishes
outside the range $x_{1}>x_{2}>x_{3}.$ The eigenvalues $B_{l}^2$ are the
same as in Eq.(21) with $f_{1}=0$, so
$$B_{l}=3(l+a+{1\over{2}})\;.\eqno(54)$$
Similarly, for the continuum states, the scattering problem for the radial
Coulomb part may be solved in the standard fashion (Landau and Lifshitz 1960),
with the phase shift $\delta_{l}$ given by the usual Coulomb expression,but
with $l$ replaced by $B_{l}$.

  The three-body problem in higher dimensions may also be mapped onto a
one-body problem. For example, the three-body problem in 2-space dimensions has
been reduced to the 4-dimensional hyperspherical coordinates (Kilpatrick and
Larsen 1987) of one particle.This
method has been used  to obtain the spectrum of three-anyons in
an oscillator potential (Khare and McCabe 1991, Law et al.\ 1992).The technique
of supersymmetric quantum mechanics, so aptly adapted for the one-dimensional
three-body problem, has not been generalised for higher dimensions
, and hence algebraic solutions  cannot be obtained  in such cases.
\vs

{\bf Acknowledgements}
\vs

This research was partially  supported by a research grant from NSERC (Canada).
R.K.B. would like to  thank the Institute of Physics and A.K. the department of
Astronomy and Physics, McMaster University, for their hospitality.
\vfill
\eject

\nd{\bf References}

\item{}
Brack M, Bhaduri R K, Law J and Murthy M V N 1993 {\it Phys.Rev.Lett.} {\bf 70}
568
\item{}
Calogero F 1969 {\it Jour.Math.Phys.} {\bf 10} 2191
\item{}
Calogero F 1971 {\it Jour.Math.phys.} {\bf 12} 419
\item{}
Calogero F and Marchioro C 1974 {\it Jour.Math.Phys.} {\bf 15} 1425
\item{}
Dutt R, Khare A and Sukhatme U P 1988 {\it Am.Jour.Phys.} {\bf 56} 163
\item{}
Erdelyi A 1955 {\it Higher Transcendental Functions} Vol. 3 (Bateman
Manuscript Project, McGraw-Hill,New York) 134
\item{}
Frahm H 1993 {\it Jour.Phys.} {\bf A26} L473
\item{}
Gandenshtein L 1983 {JETP Lett.} {\bf 38} 356
\item{}
Gutzwiller M 1990  {\it Chaos in Classical and Quantum Mechanics} (Springer
Verlag, New York)
\item{}
Haldane F D M 1988 {\it Phys.Rev.Lett.} {\bf 60} 635
\item{}
Henon M and Heiles 1964 {\it Astron.Jour.} {\bf 69} 73
\item{}
Khare A and Sukhamte U P 1991 {\it Physics News (India )} {\bf 22} 35
\item{}
Khare A and McCabe J 1991 {\it Phys.Lett.} {\bf B269} 330
\item{}
Kilpatrick J E and Larsen S Y 1987 {\it Few Body Syst.} {\bf 3} 75
\item{}
Landau L and Lifshitz E M 1960 {\it Mechanics} (Pergammon Press, Oxford )
\item{}
Law J, Suzuki A and Bhaduri R K 1992 {\it Phys.Rev.} {\bf A46} 4693
\item{}
Levai G 1989 {\it Jour.Phys.} {\bf A22} 689
\item{}
Marchioro C 1969 {\it Jour.Math.Phys.} {\bf 11} 2193
\item{}
Olshanetsky M A and Perelomov A M 1981 {\it Phys.Rep.} {\bf 71} 314 ;
1983 {\it Phys.Rep} {\bf 94} 6
\item{}
Polychronakos A P 1992 {\it Phys.Rev.Lett.} {\bf 69} 703
\item{}
Shastry B S 1988 {\it Phys.Rev.Lett.} {\bf 60} 639
\item{}
Sutherland B 1971 {\it Jour.Math.Phys.} {\bf 12} 246
\item{}
Wolfes J 1974 {\it Jour.Math.Phys.} {\bf 15} 1420

\vfill
\eject
\end